\documentstyle [12pt,epsf] {article}
\begin{document}
\begin{titlepage}
\title
{\Large \bf The odderon and spin-dependence of high energy proton-proton
scattering.}

\author{{\small \bf E. Leader} \\ {\small \it Dept. of Physics and Astronomy,
The Vrije Universiteit,} \\ {\small \it Amsterdam, The
Netherlands}\thanks{Permanent address: Birkbeck College, University of
London, U.K.}
\and {\small \bf T.~L. Trueman} \\ {\small \it Physics Department, Brookhaven
National Laboratory,} \\ {\small \it Upton, N.Y. 11973, U.S.A.}
\thanks{This manuscript has been authored under contract number
DE-AC02-98CH10886 with the U.S. Department of Energy. Accordingly, the U.S.
Government retains a non-exclusive, royalty-free license to publish or
reproduce the published form of this contributions, or allow others to do so,
for U.S. Government purposes.}}
\date{\small July 20, 1999}
\maketitle
\begin{abstract}
The sensitivity of the spin dependence of high energy $pp$ scattering,
particularly the asymmetry $A_{NN}$, to the odderon is demonstrated. Several
possible ways of determining the spin dependence of the odderon coupling
from small-$t$ data are presented.
\end{abstract}
\end{titlepage}
The odderon is a latecomer to the family of Regge poles and, to
date, there is not any firm experimental evidence for it. It is the putative
negative charge conjugation partner to the Pomeron, the dominant Regge
singularity at high energy. The exact nature of the Pomeron is even now not
well understood. Two aspects are virtually certain, almost by definition:
(1) it is a singularity, no doubt more complicated than a simple pole, in the
$t$-channel angular momentum plane that lies at $J=1$ when the momentum
transfer $t=0$, and (2) it has charge conjugation $C=+1$, signature
$(-1)^J=+1$ and isospin $I=0$. The odderon, also by definition, will lie
at or a little below $J=1$ at $t=0$. It too has $I=0$ but $C=(-1)^J=-1$. The
possiblity of such a Reggeon was first recognized by \cite{Lukaszik} and its
properties and implications have been extensively explored by
\cite{Joynson},\cite{Gauron},\cite{Ryskin} and \cite{D&L.O}. The work of
Lipatov and his collaborators
\cite{Lipatov} on the Pomeron in QCD strongly suggests that the odderon
exists on equal footing with the Pomeron \cite{LipatovO}. The QCD Pomeron is
generated by the exchange of two Reggeized-gluons in a $C=1$,
colorless state while the odderon is generated by three Reggeized-gluons in
a $C=-1$, colorless state. The QCD calculations yield a Pomeron intercept
slightly above 1 and an odderon slightly below 1. We know from unitarity
that ultimately the Pomeron intercept will lie at (or below) 1 in order
to satisfy the Froissart bound; we do not know quantitatively what such
effects will do to the odderon. (We do know that it cannot ultimately lie
{\it above} the Pomeron in order for both the $pp$ and $\bar{p}p$  total
cross-sections to be positive.) In the following we shall
simply assume thay both singularities are very close to 1. At RHIC energies
the effective intercepts may even be slightly above 1.  

The most clear-cut implication of the existence of the odderon is that it
would lead to asymptotically different amplitudes for the scattering of a
particle and its anti-particle off the same target. This means that the
total cross-sections and the differential cross-sections for,
say, $pp$ and $\bar{p} p$ scattering at high energy will remain different as
$\sqrt{s}$, the total center-of mass energy, increases; in the absence of an
odderon they would become the same, roughly as $1/\sqrt{s}$. Unfortunately,
a decisive test of this feature is not possible because of the absence of
data at the same energy for the two cases. There are suggestions that the
odderon might be important because the difference between the $pp$ and
$\bar{p}p$ differential cross-sections in the dip region appears to persist
as the energy grows 
\cite{GNL, Contogouris}. At the same time fits to
$\sigma_{\rm tot}$ and $\rho(t=0)$, the ratio of real to imaginary
parts of the forward, helicity-diagonal amplitudes, over a wide energy
range for both
$pp$ and $\bar{p}p$ leaves little room for the odderon at $t=0$
\cite{D&L.F,Block}. Recently new methods for observing the odderon in $\pi
p \to \rho n/p$ \cite{Contogouris2}, in pseudoscalar production
\cite{Nachtmann} or charm versus anti-charm jets
\cite{Brodsky} in $ep$ collsions have been proposed.

Spin-dependence of high energy
proton-proton elastic scattering provides a new and sensitive tool to search
for the odderon at small $t$. The reason for this is that the asymptotic
phase of the scattering amplitude is closely tied to the
$C=(-1)^J$ of the exchanged system; thus, in leading order, if the Pomeron
and odderon have the same asymptotic behaviour, up to logs, then
they are out of phase by $90 ^{\circ}$ \cite{Eden}. This phase condition is well-established and can
be arrived at in several ways; the most direct is to note that a Regge
singularity at $J=\alpha(t)$ in a positive signature amplitude has the
behaviour $(s^{\alpha(t)} + (-s)^{\alpha(t)}))/\sin{\pi \alpha(t)}$ while 
for negative signature it is $(s^{\alpha(t)}-(-s)^{\alpha(t)}))/\sin{\pi
\alpha(t)}$; these are each to be multiplied by functions of $t$ which real
analyticity requires to be real  in the $s$-channel physical region. Spin
dependent asymmetries depend on various real and imaginary parts of products
of amplitudes and so the odderon can dominate some asymmetries to
which the Pomeron cannot contribute. The objective of this short note is to
point out some asymmetries which might be especially sensitive to the
presence of the odderon. 

The most promising asymmetry for this purpose is the double
transverse-spin asymmetry $A_{NN}$ which will be measured in the new RHIC
spin program \cite{RHIC, Guryn};  
\begin{equation}
A_{NN} \frac{d \sigma}{dt}= \frac{4 \pi}{s^2} \{2 |\phi_5|^2 + \rm{Re}
(\phi_1^* \phi_2 - \phi_3^* \phi_4) \}.
\end{equation}
As shown by the methods in \cite{B.}, the shape of the small-$t$ dependence
of this quantity determines separately the real and imaginary parts of the
double-helicity flip $pp$ amplitude $\phi_2$. $\phi_1$ and $\phi_3$ are the
two non-flip amplitudes and $\phi_5$ denotes the single-flip
amplitude. ($\phi_4$ denotes the double-flip amplitude which vanishes by
angular momentum conservation as
$t\to 0$. It will be disregarded here.) The notation $\phi_{\pm} = (\phi_1
\pm \phi_3)/2$ is frequently used. Due to the interference between the
one-photon exchange and the strong, QCD amplitude,
$A_{NN} d\sigma/dt$ has a pole at
$t=0$. The coefficient of this pole is proportional to $\alpha \, \rm{Re} 
(\phi_2)$. As $t \to 0$ after the pole is extracted the remainder is
proportional to
$\rho \,
\rm{Re}(\phi_2) +
\rm{Im}(\phi_2)$. (This formula assumes only that the two non-flip forward
amplitudes $\phi_1$ and $\phi_3$ are equal. The quantum numbers of both the
Pomeron and the odderon are such that this is so, though lower lying
trajectories such as the $a_1$ could contribute to their difference but
should be quite negligible at RHIC energies\cite{B.}.) Because of the
singularity these terms are of comparable size for $|t|$ between
$10^{-3}$ and $10^{-2}$. The part coming from the Coulomb enhancement,
proportional to $\alpha \, \rm{Re}  (\phi_2)$, gives a characteristic peak
in $A_{NN}$ near
$t=-3 \times 10^{-3}$, while the purely strong interference between $\phi_1$
and $\phi_2$ is virtually constant in the small $|t|$ region.   This is
completely analogous to the so-called CNI peak in $A_N$ which arises
from the interference of the one-photon exchange contribution to $\phi_5$
with the imaginary part of $\phi_+$. Since 
the odderon contribution is nearly real---exactly real if it
is a simple pole at $J=1$---it will be enhanced by the CNI effect. 

This effect is illustrated in Figure 1 where curves for
$A_{NN}$ are given for three cases. The case in point (``pure odderon")
shows the peak resulting from a 5\% odderon contribution; precisely, 
$\phi_2 = 0.05 \,i \, \phi_1$. This magnitude is chosen because it gives a
value for $A_{NN}$ which is roughly at the limit of the early RHIC
experiments \cite{Guryn}. For comparison, we show a ``pure Pomeron" of the
same magnitude but $90 \deg $ out of phase: $\phi_2 = 0.05 \,\phi_1 $. The
shape is quite distinct. Finally an ``equal mixture", $\phi_2 = (0.05 \,i \,
+ 0.05 )\,\phi_1$ is shown. (In all of these cases $\phi_1$
is taken to have a $\rho$-value of 0.13.) Evidently, the odderon should be
detectable if it is this large. Since we do not know how large the odderon
double-helicity flip coupling is, or if it exists at all, we cannot predict
how large this effect will be. This illustrates how small a coupling we
can hope to learn about in the not-too-distant future.
\begin{figure}[h]
\centerline{\epsfbox{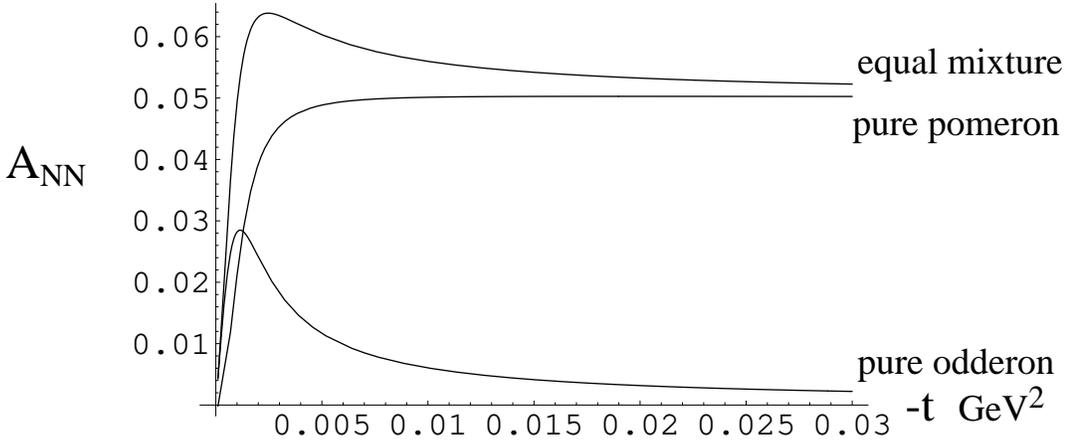}}
\medskip
\caption{\sl This illustrates the enhancement of the odderon
contribution to $A_{NN}$ due to interference with the one-photon exchange.
The three curves correspond to
$\phi_2/\phi_+ =0.05 \, i$ (pure odderon), $\phi_2/\phi_+ =
0.05$ (pure Pomeron) and $\phi_2/\phi_+ = 0.05(1 + i)$ (equal
mixture).The ``pure odderon" curve is typical of the level of sensitivity
expected for the RHIC pp2pp experiment \cite{Guryn}.}
\end{figure}

Because the Pomeron is certainly not a simple pole at $J=1$ \cite{Block,
D&L} the Pomeron will contribute a
small piece to the real part to the amplitude $\phi_2$.  Correspondingly
the odderon will contribute a small piece to the imaginary part. To have a
framework for discussing the corrections required by these pieces, we follow
\cite{B.} and write for $t \to 0$,
\begin{equation}
\frac{t}{\sigma_{\rm tot}} \, A_{NN} \, \frac{d\sigma}{dt} = 
\alpha \, a_{NN} +
\frac{\sigma_{\rm tot}}{8 \pi} \, b_{NN}
\, t +\ldots ,
\end{equation}
which separates the Coulomb enhanced piece into $a_{NN}$ and the purely
strong piece into $b_{NN}$. We disregard $\phi_5$ because it does not enter
our consideration and we assume that $\phi_-$ can be
neglected as mentioned earlier.  Then the expressions for $a_{NN}$ and
$b_{NN}$ are:
\begin{equation}
a_{NN} = \frac{{\rm Re}(\phi_2)}{2\ {\rm Im}(\phi_+)},
\end{equation}
and
\begin{equation} 
b_{NN} =\rho  \, a_{NN}+ \frac{{\rm Im}(\phi_2)}{2\ {\rm Im}(\phi_+)},
\end{equation}
where
\begin{equation}
2\ {\rm Im}(\phi_+)= \frac{s}{4 \pi} \sigma_{\rm tot}.
\end{equation}

For this discussion we will consider explicitly only the
dominant Pomeron and the odderon. We will allow the two contributions to
have slightly different energy dependence but will assume that the energy
dependence of the contributions to $\phi_1$ and
$\phi_2$ are the same so that the phases of the Pomeron piece and of the
odderon piece are the same in both amplitudes. This may not be exactly true
and may need to be corrected for, but it should not change things in an
important way.

So we will write the amplitudes $\phi_+ = (\phi_1 + \phi_3 )/2$,
\begin{eqnarray}
\phi_+ &=& A_+^P \,  e^{i\delta_P} + A_+^O \,  e^{i\delta_O}, \nonumber \\
\phi_2 &=& A_2^P \, e^{i\delta_P} + A_2^O \, e^{i\delta_O},
\end{eqnarray}
with $\delta_P \approx \delta_O + \pi/2$.
The $A's$ are real functions of $s$. Then from Eqs.\ (3-5)
\begin{equation}
A_2^P \cos{\delta_P} + A_2^O \cos{\delta_O} = \frac{s \sigma_{\rm tot}}{4
\pi}\, a_{NN},
\end{equation}
and
\begin{equation}
A_2^P \sin{\delta_P} + A_2^O \sin{\delta_O} = \frac{s \sigma_{\rm tot}}{4
\pi}\, (b_{NN} - \rho \, a_{NN}).
\end{equation}
We also have
\begin{eqnarray}
\rho & = &\frac{A_+^P \, \cos{\delta_P} + A_+^O \,
\cos{\delta_O}}{A_+^P
\sin{\delta_P + A_+^O \sin{\delta_O}}} \nonumber \\
 & \approx & \cot{\delta_P} + \frac{A_+^O \, \cos{\delta_O}}{A_+^P \,
\sin{\delta_P}},
\end{eqnarray}
since the magnitude of the non-flip odderon amplitude is less than a few
percent of the Pomeron \cite{D&L.F, Block} and in addition one expects that
$\sin{\delta_O} \approx \rho$ so the neglected term is tiny.

Note that the cross-section difference for parallel and anti-parallel
transverse spins is given by
\begin{eqnarray}
\sigma_{\rm T} & = & -\frac{8 \pi}{s} \rm{Im}(\phi_2) \nonumber \\
& = & -\frac{8 \pi}{s} (A_2^P \, \sin{\delta_P} + A_2^O \, \sin{\delta_O}) 
\end{eqnarray}
and so contains no additional information. However, it can be used as a
consistency check on the measurement of $a_{NN}$ and $b_{NN}$ since from
Eq.\ (8)
\begin{equation}
\rho \, a_{NN} - b_{NN} = \frac{\sigma_T}{2 \sigma_{\rm tot}}.
\end{equation}

With knowledge of the energy dependence of the Pomeron and the odderon,
either from theory, a model or data, one can separately determine the
phases; thus if they are simple poles behaving as $s^{\alpha_P}$ \cite{D&L}
and $s^{\alpha_O}$, respectively, their phases will be constants given by
$\sin{\delta_P} = \sin{(\pi \, \alpha_P /2)}$ and $\sin{\delta_O} =
\cos{(\pi \, \alpha_O /2)}$. Alternatively, in the asymptotic region where a
description in terms of the Froissaron and the maximal odderon
\cite{Joynson, Froissaron} is valid then $\cot{\delta_P} = \pi /\log{s}$ and
$\tan{\delta_O} = \pi /\log{s}$. Obviously, more complex behaviours are
possible; so, e.g., one must correct for contributions from lower lying
trajectories. The important point is that, because the Pomeron and the
odderon have different signature $(-1)^J$, one can determine their
magnitudes from $pp$ data without needing to use ${\bar p} p$ data.
Explicitly
\begin{equation}
A_2^O \sin{(\delta_P -\delta_O)} = \frac{s \sigma_{\rm tot}}{4 \pi} \left\{
(1 + \rho \cot{\delta_P} )\, a_{NN} - \cot{\delta_P} \, b_{NN} \right\}.
\end{equation}
Then if the
odderon phase (or energy dependence) is assumed to be known, this equation
fixes
$A_2^O$ and, via Eq.(7), determines the Pomeron double-flip amplitude
$A_2^P$.

Even without knowledge of the phases it may be possible to identify effects
of the odderon through the spin-dependence. Thus from Eqs.(7) and (8)
one sees that, in the absence of any odderon couplings,
\begin{equation}
a_{NN} = \rho \ b_{NN}/(1+\rho^2) \approx \rho \ b_{NN}.
\end{equation}
If this equality is  not true, then one can conclude that the
odderon is present in $A_{NN}$ (though the converse is not true) and can
attempt to extract more specific information from Eqs.\,(7) and (8).
Evidently, one cannot extract in a model- independent way the two odderon
amplitudes and the Pomeron double-flip amplitude from this limited number of
measurements. However, rather plausible assumptions may enable one to learn
something interesting here.

For example, it seems reasonable to suppose that the odderon intercept is
close enough to 1 that $|\sin{\delta_O}|$ is of the order of, or less than
$\rho$, as we have already done. If, in addition, we assume that the odderon
amplitudes are both, in magnitude, less than about 10\% of the Pomeron
amplitudes, then to lowest order in these small quantities we can learn that
to a very good approximation,
\begin{equation}
b_{NN} = \frac{1}{2} \left\{\frac{A_2^P}{A_+^P} \right\}.
\end{equation}
This last gives us directly an experimental determination of the
double-flip amplitude for Pomeron exchange and is insensitive to the
odderon. Next, in this approximation
\begin{equation}
\frac{s \sigma_{\rm tot}}{4 \pi}(a_{NN} - \rho \, b_{NN}) = A_O^+
\cos{\delta_O} \left(\frac{A_2^O}{A_+^O} - \frac{A_2^P}{A_+^P} \right).
\end{equation}
The odderon enters here in several ways; the most notable thing is that if
the spin structure of the Pomeron and the odderon are the same
\begin{equation}
\frac{A_2^O}{A_+^O} = \frac{A_2^P}{A_+^P},
\end{equation}
then the
term involving the odderon directly drops out and one learns the spin
structure of the odderon coupling but nothing about the magnitude beyond
that contained in
$\rho$. Model calculations by Ryskin \cite{Ryskin} suggest that this may be
nearly so. Clearly, this measurement will be most interesting if the spin
dependence of the odderon coupling is very different from that of the
Pomeron, in particular if its flip to non-flip ratio is large, as it is for
some ordinary Regge poles.

One should note, of course, that the RHIC
$pp$ program will give data for $\rho$ in an energy range which
overlaps existing $\bar{p} p$ data and one can use
\begin{equation}
\frac{A_+^O}{A_+^P} \approx (\rho(pp) - \rho({\bar p}p))/2
\end{equation}
to determine $A_+^O$ in a model-independent way. With this in hand Eq.\ (14)
and Eq.\ (15) or (16) will yield the remaining amplitudes $A_2^P$ and
$A_2^O$.

We close with a couple of related observations: (1) The $pp$ single-spin 
asymmetry $A_N$ has the well-known Coulomb enhanced peak, the height of
which depends on the imaginary part of the amplitude $\phi_5$; for $|t|$
greater than about $10^{-2}$ the purely strong interference will dominate
if there is a significant phase difference between $\phi_5$ and $\phi_1$
\cite{B.}. If both amplitudes have the same asymptotic behaviour they will be
in phase unless the odderon couples to one or the other, and so a measurement
of $A_N$ above the CNI peak which does not decrease rapidly with energy is
another signal for odderon coupling. See however \cite{Ryskin}. (2) A very
similar discussion could be carried through for the double longitudinal spin
asymmetry $A_{LL}$ with
$\phi_-$ replacing $\phi_2$. Since the odderon has the wrong quantum numbers
to couple to this amplitude ---it requires $(-1)^J = - C$ ---a non-zero value
asymptotically for $a_{LL}$ , which is proportional to ${\rm Re}(\phi_-)$,
would be a strong indication for yet another Regge singularity near $J=1$.
This is not subject to corrections coming from the Pomeron since it cannot
couple to $\phi_-$ at all. We are not aware of any theoretical argument for
such a singularity; thus, the observation of such an asymmetry would be
extremely interesting.

{\bf Acknowledgements} E.L.is grateful to the Foundation for Fundamental
Research on Matter (FOM) and the Dutch Organization for Scientific
Research (NWO) for support.

\end{document}